\documentclass[a4paper,12pt]{article}
\usepackage[utf8x]{inputenc}
\usepackage{amssymb,amsmath}
\usepackage{cite}

\newtheorem{Theorem}{Theorem}[subsection]

\newtheorem{lem}[Theorem]{Lemma}

\newtheorem{df}[Theorem]{Definition}

\newtheorem{rem}[Theorem]{Remark}

\newcommand{\comment}[1]{}

\def\phi{\varphi}

\newcommand\Z{\mbox{$\mathbb Z$}}
\newcommand\M{\mbox{$\mathbb M$}}
\newcommand\N{\mbox{$\mathbb N$}}

\newcommand\F{\mbox{$\mathbb F$}}
\newcommand\A{\mbox{$\mathcal A$}}

\newcommand\G{\mbox{$\mathbb G$}}

\newcommand\V{\mbox{$\mathcal{V}$}}

\newcommand{\marginlabel}[1]%
{\mbox{}\marginpar{\it{\raggedleft\hspace{0pt}#1}}}

\newcommand\pr{\mathop{\bf {Pr}}}

\def\N{{\mathbb{N}}}
\def\L{\mathbb{L}}

\title{Public-Key Cryptography Based on Modular Lattices}
\author{Franti\v sek Polach}

\begin{document}

\maketitle

\begin{abstract}
We present an approach to generalization of practical Identity-Based Encryption scheme of ~\cite{BonFra}. In particular we show how the protocol could be used on finite modular lattices and as a special case on vector spaces over finite field.
The original proof of security for this protocol does not hold in this general algebraic structure, thus this is still a work in progress.
\end{abstract}

\section{Introduction}

The goal of this work is to investigate the possibility of generalization of practical Identity-Based Encryption scheme of ~\cite{BonFra} into a different algebraic structure. Specifically we use finite modular lattices instead of cyclic groups and replace the original pairing on elliptic curves with a special pairing on modular lattices.

\section{Preliminaries}
Here we introduce the necessary formalism.
\subsection{Security Model for Identity-Based Encryption}
We review the standard security model for Identity-Based Encryption as can be found in ~\cite{BonFra, Gentry06}.\\

An IBE scheme consists of four randomized algorithms: \emph{Setup, KeyGen, Encrypt}, and
\emph{Decrypt}. 
\emph{Setup} sets the Private Key Generator's (PKG) parameters \emph{params} and a master key \emph{master-key}. 
\emph{KeyGen} is a probabilistic algorithm that generates a private key for an identity using master-key. 
\emph{Encrypt} encrypts a message, taking an identity and params as input, and outputs a ciphertext. 
\emph{Decrypt} decrypts a ciphertext for an identity using a private key for that identity.\\

Boneh and Franklin define \bf{IND-ID-CCA security }\normalfont (indistinguishability under adaptive identity and adaptive chosen ciphertext attack) via the following game.
\medskip

\begin{minipage}{6in}
 
\bf{Setup}: \normalfont
 The challenger runs \emph{Setup}, and sends params to the adversary while keeping
the master key to itself.\\

\bf{Phase 1}: \normalfont
The adversary issues queries $q_1, \dots , q_m$ where
$q_i$ is one of the following:
\begin{itemize}
\item Key-extraction query $\langle ID_i \rangle$: the challenger runs \emph{KeyGen} on $ID_i$ and forwards
the resulting private key to the adversary.
\item Decryption query $\langle ID_i, C_i \rangle$: the challenger runs \emph{KeyGen} on $ID_i$, decrypts $C_i$ with the resulting private key, and sends the result to the adversary.
\end{itemize}
Adversary can make these queries adaptively, i.e., any query may depend on the
previous queries as well as their answers.\\

\bf{Challenge}: \normalfont
 The adversary submits two equal length plaintexts $M_0, M_1 \in \M$ and an identity
$ID$. Obviously, $ID$ must not have appeared in any key generation query in Phase 1. The
challenger selects uniformly at random a bit $b \in \{0, 1\}$, obtains a ciphertext $C = Encrypt(params, ID, M_b)$,
and sends $C$ to the adversary as its challenge ciphertext.\\

\bf{Phase 2}: \normalfont
This is identical to Phase 1, except that the adversary may not request
a private key for $ID$ or the decryption of $(ID,C)$.\\

\bf{Guess}: \normalfont
The adversary outputs a guess $b' \in \{0, 1\}$ for $b$. The adversary wins if $b = b'$.\\

The advantage of the adversary in attacking the IBE scheme is defined as:
\begin{eqnarray}
Adv = |\Pr[(b = b')] − 1/2|
\end{eqnarray}
\end{minipage}
\medskip

We call the adversary in the above game an IND-ID-CCA adversary.\\

\begin{df}
An IBE system is $(t, q_{ID}, q_C, \epsilon)$ IND-ID-CCA secure if all $t$-time
IND-ID-CCA adversaries making at most $q_{ID}$ private key queries and at most $q_C$
chosen ciphertext queries have advantage at most $\epsilon$ in winning the above game.
\end{df}

\bf{IND-ID-CPA security }\normalfont is defined similarly, only with the restriction that the adversary cannot make decryption queries.

\begin{df}
An IBE system is $(t, q_{ID}, \epsilon)$ IND-ID-CPA secure if it is $(t, q_{ID}, 0, \epsilon)$ IND-ID-CCA secure.
\end{df}

\subsection{Bilinear Maps and Pairings}

We will generalize the standard notion of a bilinear map ~\cite{BonFra, BonBoy04, Gentry06}. The standard setting is following:

\begin{itemize}
 \item $\G$ and $\G_T$ are two (multiplicative) cyclic groups of prime order $p$;
 \item $g$ is a generator of $\G$.
\end{itemize}

Let $\G$ and $\G_T$ be two groups as above. An (admissible) \textit{bilinear map } is a map $e : \G \times \G \rightarrow \G_T$
with the following properties:
\begin{itemize}
 \item Bilinearity: for all $u, v \in \G$ and $a, b \in \Z$, we have $e(u^a, v^b) = e(u, v)^{ab}$.
 \item Non-degeneracy: $e(g, g) \ne 1$.
 \item Computability: there is an efficient algorithm to compute  $e(u, v)$ for any $u,v \in \G$
\end{itemize}

We say that $\G$ is a bilinear group if the group action in $\G$ can be computed
efficiently and there exists a group $G_T$ and an efficiently computable bilinear
map $e : \G \times \G \rightarrow \G_T$ as above. Note that $e(, )$ is symmetric since $e(g^a, g^b) =
e(g, g)^{ab} = e(g^b, g^a)$.\\

We will call the generalization of the bilinear map a pairing:

\begin{df}
Let $X, Y$ be finite sets and $A$ be a semigroup acting on $X$ from the left. 
Then a mapping $e: X \times X \rightarrow Y$ is called a \emph{pairing} on $X$ iff $e$ is bilinear, that means:
$$e(ax_1,x_2) = e(x_1,ax_2), \text{ for all } x_1, x_2 \in X \text{ and } a \in A.$$

\end{df}
Note also that we will use the bilinear property in both coordinates:
\[
 e(ax_1,bx_2)=e(x_1,abx_2)=e(x_1,b(ax_2))=e(bx_1,ax_2). 
\]

\subsection{Complexity Assumptions}

Let $X, Y$ be finite sets and $A$ be a semigroup acting on $X$ and $Y$ from the left. We will assume the following problems are hard:
\begin{itemize}
 \item Discrete Log problem (DLP): $p \in X$, for given $(p,ap)$ determine $a$.
 \item \bf{Bilinear Diffie-Hellman (BDH) }\normalfont in $\langle X, Y, e \rangle$ ~\cite{BonFra}: 
for given $(x, y, ax, bx)$ find $e(ax, by)$,
for $x,y \in X$.
\end{itemize}

An algorithm $\A$ has advantage $\epsilon$ in solving BDH in $\langle X, Y, e \rangle$ if
\[
 \pr [\A(x, y, ax, bx) = e(ax, by) ] \geq \epsilon
\]
where the probability is over the random choice of $a,b \in A$, the random choice of $x, y \in X$, and the random
bits of $\A$.\\

\medskip

\bf{BDH Assumtion. }\normalfont \\
We say that the BDH problem is $(t, \epsilon)$-hard in $X$ if no $t$-time algorithm can solve BDH problem with advantage
at least $\epsilon$.

\bf{Hardness of BDH. }\normalfont \\
The BDH problem in $\langle X, Y, e \rangle$ is no harder than the Computational Diffie-Hellman problem (CDH) in $X$ or $Y$. 
The converse is still an open problem: is an algorithm for BDH sufficient for solving CDH?.\\
The best known algorithm for BDH is to solve DLP in either $X$ or $Y$.

\medskip

\subsection{Brief modular lattice theory}

Now we introduce some basic notions of lattice theory, a good reference is the book ~\cite{latt_book}.

\begin{df}
A poset $(L,\leq)$ is called a \emph{lattice}, if for every finite subset $A \subset L$ there exists a \emph{join} 
(least upper bound) $\bigvee A$ and a \emph{meet} (greatest lower bound) $\bigwedge A$ in $L$.\\
For a finite $L$, we define the \emph{least} and the \emph{greatest} element of $L$ respectively as $O:=\bigwedge L$ and $I:=\bigvee L$.\\
For $a,b \in L$, an \emph{interval} is 
$$[a,b] := \{x \in L | a \leq x \leq b\}.$$
\end{df}

\medskip

An equivalent universal algebraic definition of lattice is
\begin{df}
An algebra $(L, \land, \lor)$ is called a \emph{lattice} if $L$ is a nonempty set,
$\land$ and $\lor$ are binary operations on $L$, both $\land$ and $\lor$ are idempotent, commutative, and associative,
and they satisfy the absorption law.
\end{df}

\medskip

A note on notation:\\
in the rest of the text we will be using symbols $+$ and $\cdot$ instead of more common $ \vee,\; \wedge $ for the two binary operations on a lattice, respectively.

\begin{df}
A lattice $\L$ is called \emph{modular} if for all $a,b,c \in L$ the following holds:
\begin{equation}
 a \leq c \Rightarrow (a + b) \cdot c = a + b \cdot c.
\end{equation}
\end{df}

An example is a normal subgroup lattice of a group is modular.\\

\begin{lem}
For a lattice $L$, the following is equivalent:
\begin{itemize}
 \item $L$ is modular;
 \item $L$ does not contain the lattice $N_5$ as a sublattice;
 \item for all $a,b,c,d \in L$
$$d \leq b \Rightarrow (a \cdot b + c) \cdot d = (a + c \cdot b) \cdot d.$$
\end{itemize}
\end{lem}

\medskip

\begin{df}
A lattice is said to be \emph{distributive} if it satisfies for all $x, y, z \in L$ either (and therefore both) of the distributive laws:
\begin{itemize}
\item $x \cdot (y + z) = (x \cdot y) + (x \cdot z)$
\item $x + (y \cdot z) = (x + y) \cdot (x + z).$
\end{itemize}
\end{df}
\medskip

The following is an easy observation.

\begin{lem}
Every distributive lattice is modular.
\end{lem}

Examples of distributive lattices include Boolean lattices, totally ordered sets, and the subgroup lattices of locally cyclic groups.

\begin{df}
A \emph{complement} of $a$ in a lattice $\L$ with $O$ and $I$ is an element $b \in L$ such that
$$a \cdot b = O \; \text{ and } \; a + b = I. $$
A bounded lattice $\L$ is \emph{complemented} if all its elements have complements.
\end{df}

\begin{rem}
Complements may not exist. If $L$ is a non-trivial chain, then no element (other than $O$ and $I$) 
has a complement. This also shows that if $a$ is a complement of a non-trivial element $b$, then $a$ and $b$ form 
an antichain.
\end{rem}

In a complemented lattice, there may be more than one complement corresponding to each element.

\begin{df}
 Two elements are said to be \emph{related}, (or \emph{perspective}) if they have a common complement.
\end{df}

\begin{rem}
If a complemented lattice $L$ is a distributive lattice, then $L$ is uniquely complemented (in fact, a Boolean lattice).

\end{rem}

\section{The generalized IBE protocol}

In the following we will present an approach to a generalization of an IBE protocol by Boneh and Franklin ~\cite{BonFra}.

\subsection{The Boneh-Franklin protocol on modular lattices}
\begin{df}
Let $\L$ be a modular lattice and $d \in L$ is fixed. Then the semigroup $A:= [d,1]$ acts on $L$ and the mapping
\begin{equation}
 e_d : L \times L \rightarrow [0,d], \; (x, y) \mapsto d \cdot (x + y)
\end{equation}
is said to be a \emph{lattice pairing}.
\end{df}

It is easy to see that the lattice pairing is indeed a pairing.\\

\medskip

We assume the pairing $e$ and the $A$-action are both nondegenerate and efficiently computable.\\

\begin{df}
Let $\F_q$ be the field with $q$ elements and $n \in \N$. The \emph{projective space} (geometry) of dimension $n-1$ and order $q$
is the lattice
\begin{equation}
 L(\F_q^{n}) := \{S \subset \F_q^{n} | S \text{ is a subspace}\}.
\end{equation}
\end{df}

Subspaces of dimension $1, 2, \dots, n-1$ are referred to as points, lines, $\dots$, hyperplanes in this geometry.\\

The \emph{Gaussian coefficients} determine the number of subspaces of a given dimension $k$. The formula is
\begin{equation}
 {n \brack k}_q := \frac{(1-q^n)(1-q^{n-1}) \dots (1-q^{n+1-k})}{(1-q^k)(1-q^{k-1}) \dots (1-q)}. 
\end{equation}

\medskip

Just for the sake of clarity we take a small example of a lattice $\L := L(\F_q^{5})$ for some $q$ and show the use of the protocol for ID-based encryption devised by Boneh and Franklin in 2001. In this case we can use standard geometrical names for our objects.
The one, two, three and four - dimensional subspaces of the projective space $L(\F_q^{5})$ are called points, lines, planes and hyperplanes, respectively.\\
There are two users with some identities and a trusted authority (TA) issuing user's private keys in the protocol.\\

Given $\L$ and a fixed line $d \in \L$, the protocol proceeds as follows:\\

\begin{minipage}{5in}
 
\bf Setup: \normalfont
 TA chooses a plane $P \in \L$, $d \not\leq P$, a hyperplane $s \in_R \L$, $d \leq s$, $P \not\leq s$, computes a line $P_{pub} := P \cdot s$, chooses cryptographic hash functions $H_1 : \{0, 1\}^* \rightarrow \L$ and $H_2 : \L \rightarrow \{0, 1\}^n$,
where $n$ is the bit length of messages.\\
The private master key is $s$ and the global public key is $P_{pub}$.

\bf Extract: \normalfont
given a user's public $ID \in \{0, 1\}^*$, compute the user's public key $Q_{ID} = H_1(ID) \in \L$ (a plane), $d \not\leq Q_{ID}$, $Q_{ID} \not\leq s$, and the private key $S_{ID} = s \cdot Q_{ID}$.

\bf Encrypt: \normalfont
 given message M, choose a secret hyperplane $r \in_R \L$, $d \leq r$, $P \not\leq r$, $Q_{ID} \not\leq r$, and compute
 $$C=\{r \cdot P, M \oplus H_2(e_d(Q_{ID} \cdot r, P_{pub}))\}.$$

\bf Decrypt: \normalfont
 given the ciphertext $C=\{U,V\}$, recover the plaintext 
 $$M=V \oplus H_2(e_d(S_{ID},U)).$$
\end{minipage}

\medskip

To verify the correctness of the protocol just substitute for $S_{ID}, U, V$ and use the property of pairing.\\

\begin{rem}
The protocol does not work for distributive lattices, as
$$e_d (Q_{ID}r, Ps) = d(Q_{ID}r + Ps) = dQ_{ID}r + dPs = Q_{ID}d + Pd,$$
which is a public element.
\end{rem}
\medskip

We can make the following observations:
\begin{itemize}
 \item The projective line $[0, d]$ has $q+1$ (nontrivial) points.
 \item The number of lines of the plane $[d, 1]$ (choice of $r, s$) is $q^2+q+1$.
 \item The requirements $d \leq s$ and $d \leq r$ cannot be dropped as we need them for the bilinearity of $e_d$.
 \item Thus there are only $(\dim s - \dim d)$ unknown dimensions in $s$ and similarly for $r$.
 \item Therefore for practical use it is necessary to consider a projective space with much higher dimension.
 \item A good choice of a dimension of the element $d$ seems to be $n/2$ as then both 
the set of hyperplanes containing $d$ (choice of $r, s$) and the set of elements contained in $d$ (range of $e_d$) are
about the same size (and 'large' enough).
\end{itemize}

\medskip
There are several nontrivial possibilities for choosing other elements of the protocol, one such choice could be this one:
\begin{itemize}
 \item $rP, sP, rQ_{ID}, sQ_{ID}$ are lines different from $d$.
 \item $rP + sQ_{ID}$ is a hyperplane different from $r$ and $s$.
 \item The pairing $e_d (rP, sQ_{ID}) = (rP + sQ_{ID})d$ is a point.
\end{itemize}

\medskip

A general weakness of this protocol is that repeated use of the system enables a user to learn about the choice of $s$. This can be resolved by bounding the number of issued private keys for a given master key.

\medskip

In the \textbf{general case} of $\L := L(\F_q^{n})$, where $n$ is large enough, we suggest the following choices:
\begin{itemize}
	\item The element $d$ is chosen from $\lfloor n/2 \rfloor$-dimensional elements of the lattice.
	\item The element $P$ is chosen among the elements of dimensions ranging from $\lfloor n/2 \rfloor + 1$ to $n - 3$, so that it does not contain $d$ (avoids the sublattice $[d, 1]$).
	\item The secret key $s$ is selected uniformly at random from the elements of dimension $n-1$ (and possibly also of dimension $n-2$) that are contained in the sublattice $[d, 1]$.
	\item The hash function $H_1$ maps the user's ID to the lattice elements of dimensions ranging from $\lfloor n/2 \rfloor + 1$ to $n - 2$, so that it does not contain $d$ (and is not contained in $s$). But is it necessary to exclude the element $P$?
	\item The secret $r$ is selected uniformly at random from the elements of dimension $n-1$ and $n-2$ that are contained in the sublattice $[d, 1]$
	\item The pairing $e_d$ maps to the sublattice $[0, d]$ (avoiding 0 and $d$), that has height $\lfloor n/2 \rfloor$.
\end{itemize}

\medskip

What is missing: proof of hardness of BDH problem in $L(\F_q^{n})$ and investigation of feasibility of the pairing.\\
This should include
\begin{itemize}
 \item a bound on the size of $q(n)$;
 \item a good representation of modular lattices (like vector spaces).
\end{itemize}

\subsection{The special case of a vector space}

In this section we will consider a special case of a finite modular non-distributive lattice of subspaces of an $n$-dimensional vector space
 $\V := \F_q^{n}$ over finite field $\F_q$.\\

We will represent this vector space and its subspaces in a standard way, i.e. by matrices of size $m \times n$
 over $\F_q$ for some $m$. Representing operations of intersection and union of subspaces is easy too as there is 
 a basis in every vector space.\\

If we choose uniformly at random two vector subspaces $V_1$ and $V_2$ of $\V$, $dim V_1 = m'_1 < n$
 and $dim V_2 = m'_2 < n$, we would like to know the expected value of $dim (V_1 \cup V_2)$.\\
We can easily compute the expected dimension of intersection of these two random spaces
\begin{equation}
	E( dim (V_1 \cap V_2) ) = E( dim V_1 ) + E( dim V_2) - E( dim (V_1 \cup V_2) ),
\end{equation}
as the intersection and union of two vector spaces is also a vector space and expectation is a linear function.\\

Let $M_1$ and $M_2$ be two matrices created by randomly sampling $m_1$ and $m_2$ vectors of length $n$ respectively (their ranks might be smaller than $m_1$ and $m_2$). These matrices will represent the random subspaces $V_1$ and $V_2$. We are interested in the dimension of the union of these two subspaces,
 thus we join the two matrices and investigate the rank of the matrix of size $m \times n$,
 where $m=m_1 + m_2$.\\

In the same way as did Linial and Weitz in ~\cite{Linial00}, we denote the collection of $m \times n$ matrices over $F_q$ by $M_{m,n,q}$,
 and this same set with a uniform distribution is the probability space $\Omega_{m,n,q}$. The rank of matrices can
be seen as an integer-valued random variable on $\Omega_{m,n,q}$.\\

The rank distribution is well known:

\begin{lem}
Let $0 \leq r \leq min\{m,n\}$, $m \leq n$ and $M$ be a matrix from $\Omega_{m,n,q}$.

Then

\begin{equation}
	\pr(rank M = r)
	= \frac{1}{q^{(n-r)(m-r)}}
	\prod_{i=0}^{r-1}
	\frac{(1-q^{i-n})(1-q^{i-m})}{1-q^{i-r}}.
\label{eq.rank}
\end{equation}

\end{lem}

In the special case when $r = m$, the matrix must be regular, so every newly added vector is independent from the linear span of the previously added vectors.

$$\pr(rank M = m) = \prod_{i=0}^{m-1}(1 - q^{i-n}).$$

More details behind the following valuable observations can be found in Linial and Weitz ~\cite{Linial00}:
\begin{itemize}
 \item A randomly chosen $m \times n$ matrix $M$ has almost surely full rank (i.e. $min \{m,n\}$) iff $|n - m|$ is unbounded (grows to infinity with $n$).
 \item If $|n - m|$ is bounded, then $\pr(rank M \leq r) \rightarrow 0$ iff $r \leq min\{m, n\} - \omega(1)$, i.e. almost full rank is almost certain.
\end{itemize}

Here "almost surely" means that the probability grows to 1 exponentially with increasing $n$.

\medskip

The description of the B-F protocol for the special case of a vector space with some concrete choices of sizes of elements follows.\\

\begin{minipage}{6in}

\bf{Setup: }\normalfont
Given n-dimensional vector space over $F_q$, choose uniformly at random $\lceil 5n/16 \rceil$ vectors to form the parameter $d$.
For $P$ randomly choose $\lceil n/2 \rceil$ vectors, to form $S$ take the vectors of $d$ and add $\lceil 9n/16 \rceil$
 random vectors.\\

\bf{Extract: }\normalfont
For a given $ID \in \{0,1\}^*$, choose $\lceil n/2 \rceil$ vectors to form $Q_{ID}$ and secret $R$ will be formed
 in the same way as $S$. Set the private key to $S \cdot Q_{ID}$.\\

\bf{Encrypt \& Decrypt: }\normalfont
 Same as before. Analysis follows.\\
\end{minipage}
\medskip

According to the observation, with high probability all these subspaces have full dimension.\\

Then the hard problem in the protocol is:\\
 for given $(Q_{ID}, P_{pub}, P \cdot R, P)$, find $e_d(Q_{ID} \cdot R, P_{pub})$.\\

This problem is not harder than finding $R, S$ given $P, Q_{ID}, Q_{ID} \cdot R, P \cdot S$, which could be seen as a variant of the discrete logarithm problem in vector spaces. But the exact relation to the original Billinear Diffie Hellman problem on elliptic curves is unclear.\\

The dimensions of $Q_{ID} \cdot R$ and $P_{pub} = P \cdot S$ is with high probability equal to 
$\lceil 7n/8 \rceil + \lceil n/2 \rceil - n = \lceil 3n/8 \rceil$.
Thus with high probability the dimension of $Q_{ID} \cdot R + P_{pub}$ is equal to $\lceil 3n/4 \rceil$. 
And so finally the dimension of 
$e_d(Q_{ID} \cdot R, P_{pub})$ is with high probability equal to $\lceil n/16 \rceil$.\\

We can easily compute the number of spaces that contain some fixed space as a subspace due to the duality of the structure
and using Gaussian coefficients.\\

The number of vector spaces of dimension $\lceil 3n/4 \rceil$ that contain a fixed space of dimension $\lceil n/2 \rceil$
is ${\lceil n/2 \rceil \brack \lceil n/4 \rceil}_q \geq q^{n^{2}/16}$. And the number of subspaces of dimension $\lceil 3n/4 \rceil$
is ${ n \brack \lceil 3n/4 \rceil}_q \geq q^{3n^{2}/16}$.\\

\medskip
\subsection{In a search for security proof}

The reason why we cannot use the same proof technique as in ~\cite{BonFra} is simple: in a vector space we are missing the multiplicative group structure. \\

The original Boneh-Franklin proof of IND-ID-CPA security of the identity based protocol is done in two steps.\\
The first step is
to show that if an adversary can break the ID based protocol with some advantage $\epsilon$, then there is an algorithm 
that can break a standard public key protocol (called BasicPub in ~\cite{BonFra}) with advantage $\frac{\epsilon}{e(1+q_E)}$, 
where $q_E$ is the number of queries to the random oracle used
for issuing public keys. This standard protocol is similar to the ID based one, just the queries for private keys are removed.\\
The second step is to show that if an adversary has some advantage against the public key protocol (without identity queries 
for public keys) then there is an algorithm that solves BDH problem with some non-trivial advantage.\\

The first part of the proof actually shows that private key extraction queries do not help the adversary.
It does that by constructing a table that substitutes the hash function answers for these queries. But
unfortunately, it relies on the existence of inverse elements in groups which we cannot get in a vector space.\\
Concretely when we simulate the first hash function we output $b \cdot Q_{ID}$ as the public key for 
the identity on which the adversary wants to be challenged.\\
We want to decrypt $C=\{ U, V\}$, thus we compute an inverse of $b$ in the group, 
 and send the adversary the ciphertext $C'=\{ b^{-1} \cdot U, V\}$.\\
We observe that decryption of $C'$ using private key $s \cdot b \cdot Q_{ID}$ is the same as decryption of $C$ using
$s \cdot Q_{ID}$: 
\[
 e(b^{-1} \cdot U,s \cdot b \cdot Q_{ID}) = e(U,s \cdot b^{-1} \cdot b \cdot Q_{ID}) = e(U,s \cdot Q_{ID}).
\]

In the second part, we have to simulate the other hash function used in the ID-based protocol. But this function maps from
a group to $\{0,1\}^n$ and thus we can easily replace the group with the vector space, as we are not using the structural
properties of a group.\\
As IND-ID-CCA security is stronger than IND-ID-CPA security, this proof will not work for it either.

\medskip
As we do not have the advantage of multiplicative group structure in a vector space, we cannot use the standard proof technique.

\subsection{Another approach: direct product $\L^n$}      

Another approach to this problem might be to take a small (possibly the smallest) lattice  $\L$ (vector space) for which the aforementioned protocol is non-trivial and then do the direct product of sufficiently large number of copies of this lattice. This construction will give us an exponentially large secret key.

\section{Conclusions} 
We failed to properly generalize the original Boneh-Franklin protocol to the case of modular lattices. The security proof of the original protocol depends on the multiplicative group structure which we are loosing when generalising to modular lattices. One possible way to proceed might be to use the direct product $\L^n$ of $n$ small lattices $\L$ for which is the protocol non-trivial.


\bibliographystyle{alpha}
\bibliography{wheel}

\end{document}